# Synergy of the Cosmic Ray and High Energy Atmospheric Physics: particle bursts observed by arrays of particle detectors

A. Chilingarian [a, b], G. Hovsepyan [a]

[a] *A. Alikhanyan National Lab (Yerevan Physics Institute), Yerevan 0036, Armenia*
[b] *National Research Nuclear University MEPhI, Moscow 115409, Russia*

**Abstract**

The particle bursts detected on the earth's surface during thunderstorms by various particle detectors pose the question of their origin and electron acceleration mechanisms. Well-known extensive air showers (EASs) originate by galactic protons and fully-stripped nuclei can cover several square kilometers and EAS particles have a very high density around the shower axes. However, EAS comprises a very thin disc of particles (a few tens of ns), and when EAS core hits detectors with a usual time resolution ($\geq 1$ μs) will generate only one pulse in spite of possibly thousands of particles are incident.
The particle fluxes, which are registered during thunderstorms originated by the relativistic runaway electron avalanches (RREAs) initiated by free electrons accelerated in the strong atmospheric electric fields. Two oppositely directed dipoles in the thundercloud accelerate electrons in the direction to the earth's surface, and – open space. The particle bursts, which are observed by orbiting gamma observatories called terrestrial gamma flashes (TGFs), and ones registered by particle detectors located on the ground – thunderstorm ground enhancements (TGEs). Balloons and aircraft in the troposphere register gamma glows. Recently to these phenomena usually are added downward TGFs (DTGFs), intense and very short particle bursts sometimes coinciding with lightning flashes. We compare available data on enhanced particle fluxes and reveal relations between extended and short particle bursts with EAS phenomena. We also demonstrate that the neutron monitors can extend the EAS "life" up to a few milliseconds, a time comparable with DTGFs duration. The possibility to use the network of neutron monitors for high-energy cosmic ray research is also deliberated.

**Plain Language Summary** Short and extended particle bursts are registered in the space, troposphere, and the earth's surface. Coordinated monitoring of the particle fluxes, near-surface electric fields, and lightning flashes makes it possible to formulate a hypothesis on the origin of intense bursts and their relation to extensive air showers and atmospheric discharges. Analysis of the observational data and possible origination scenarios of particle bursts allows us to conclude that the bursts can be explained by the electron acceleration in the thunderous atmosphere and by gigantic showers developed in the terrestrial atmosphere by high-energy protons and fully-stripped nuclei accelerated in Galaxy and beyond.

# 1. Introduction

After the discovery of terrestrial gamma flashes (TGFs, [1]) the high-energy physics in the atmosphere (HEPA) is gaining increasing attention by both measurements with gamma-ray instruments onboard satellites and by numerous theoretical and modeling studies [2]. However, a very complicated experimental arrangement (particle detectors are located ≈500 km from the radiation source on the fast-moving satellite) and the absence of an online trigger for particles coming from the earth's direction make the identification of TGF origin a rather complicated problem. Till now, the relation of TGFs to lightning discharges is under debate and the scheme of TGF origination itself is unclear. The runaway breakdown (RB) model introduced in [3] and afterward mostly cited as relativistic runaway electron avalanche (RREA, [4,5]) did not produce gamma ray beam enough intense to describe satisfactorily the TGF observations [6]. The situation drastically changed when observation of enhanced fluxes of electrons, gamma rays, and neutrons start to be performed on the earth's surface with numerous particle detectors located just below electron-photon avalanches [7]. Registration of electron-gamma ray avalanches emerging in the electrified atmosphere made it possible to get inside in the mechanisms responsible for thunderstorm ground enhancements (TGEs, [8-11]). Simulations with GEANT4 [12] and CORSIKA codes [13] confirm that the origin of the TGEs observed on Aragats can be explained by the RB/RREA process assuming atmospheric electric fields above the critical value, which were routinely measured in direct experiments [14,15]. There is no need to introduce additional complications of the RB/RREA process, particle multiplication naturally emerges by introducing the strong electric field in the atmosphere above particle detectors. Certainly, in order to obtain the energy spectra alike measured in the experiment, several appropriate combinations of the electric field strength and its spatial extension should be examined [10]. Simulations confirm that free electrons from EASs, generate multiple electron–photon avalanches (extensive cloud showers – ECSs), which cover sizable areas on the ground and can induce surface array triggers, as EASs do. However, the density and energy of ECS particles are much smaller than EASs (see Figs 3-5 in [8] and Fig. 5 in [9]).
The facilities installed on Aragats station for the measurements of atmospheric discharges are synchronized on a nanosecond time scale with particle detectors making it possible to study the interrelation of TGEs and lightning flashes [16-19]. Multiyear measurements allow us to claim that atmospheric discharges do not originate particle fluxes, but abruptly terminate them [20]. Recently, the research groups using large arrays of particle detectors deployed for the registration of EASs, became interested in the unusual triggers (particle bursts) and their possible correlations with lightning activity. Special attention is paid to establishing a combined monitoring technique of high precision registration of particle fluxes and atmospheric discharges. Coordinated monitoring of the lightning flashes and particle bursts, as well as, the modeling of the avalanche transport in the electrified atmosphere, make it possible to formulate hypotheses on the origin of particle bursts and their relation to atmospheric discharges.

In this paper, we discuss several scenarios of the origination of particle bursts. In the second section of the paper, we demonstrate the emergence and terminating of multiple TGEs on a stormy day of 9 June 2021, and the role of atmospheric discharges in stopping TGEs.
In the third section, we show how Aragats neutron monitor (ArNM) extends the "live" of the EAS ≈100000 times, from a few tens of ns to a few ms. In the fourth chapter, we discuss the downward TGFs (DTGFs) using as examples burst registration by HAWC [21] and TA [22] experiments.

## 2. Particle fluxes and lightning discharges

In Fig. 1 we present the time series of particle fluxes, disturbances of the near-surface (NS) electric field, and distances to lightning discharges measured by large scintillation spectrometer ASNT (see details in [9]) and electric field sensor EFM-100 produced by BOLTEK company. The accuracy of the TGE duration estimation is a few seconds; of the time of lightning flash – a µs (well coinciding with worldwide lightning location network time stamp when available); the relative accuracy of the count rate measurements is less than 1%.
We can see 5 successive attempts to start a TGE that were stopped by lightning discharges. The lightning flashes are seen in the time series the NS electric field as an abrupt increase (or decrease) of the electric field lasting a few hundreds of ms continued by recovering lasting several tens of seconds. No sharp changes of the near-surface electric field have been detected during the rising phase of TGE, that is, there were no signs of any specific atmospheric discharges that possibly can initiate enhanced particle flux. When atmospheric conditions permit, the TGE duration exceeds a minute, however, the duration of 3 of 5 TGEs was less than a minute due to lightning flashes, which decrease the potential difference in the accelerating dipole below the critical value and stop the runaway process. In Table 1 we present detailed information on 5 short TGE events and on lightning flashes that terminate 4 of them. In the second column of the table, we show the TGE duration, in the third - the type of atmospheric discharge that stopped it (the type of lightning flash was determined by the method described in [18]). The TGEs were terminated by the inverted polarity -IC), normal polarity (+IC), and cloud to ground (-CG) flashes. In the third column, we show the measurements of the near-surface (NS) electric field. According to TGE initiation scenarios (see Fig. 1 in [11]) both positive and negative NS electric fields are observed during TGE. In the fourth column, we post the distance to the lightning flash estimated by the same EFM-100 sensor. The accuracy of distance measurements is ≈1.5 km. In the last column, we show the TGE significance (magnitude, strength): the percent of enhancement above background measured at fair weather before TGE. All small TGEs demonstrate peak values above 3 standard deviations.

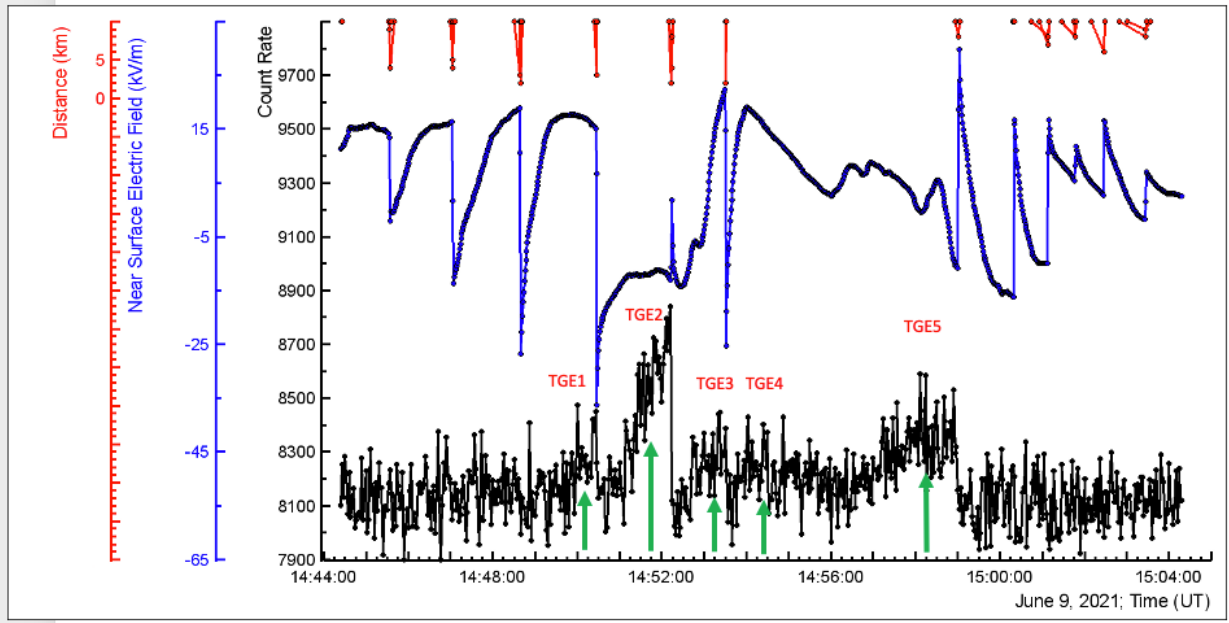

**Figure 1. One-second time series of NS electric field measurements (blue, EFM-100); of count rate of 60 cm thick and 4 m$^2$ area plastic scintillator (black, ASNT detector, energy threshold 4 MeV); by the red lines we denote the distance to the lightning flash estimated by the same EFM-100 "electric mill", by green arrows - the 5 TGEs four of which are terminated by the lightning flash.**

Considered thunderstorm illustrates a typic relation between TGEs and lightning flashes. If the minimum of the NS electric field is deeper (reaching -20 - -30 kV/m), or the duration of the positive NS field is larger (reaching 5 minutes) we can expect much more intense TGEs reaching 100% flux enhancement for the 1-minute count rate time-series, according to scenarios of TGE initiation described in [11]. For the 4 TGEs observed on June 9, 2021, the distance to the lightning flash was 1.6 – 7.5 km. We can assume, that the lightning struck nearby the place above which the potential drop (voltage) is maximum, i.e., at the distance of 7.5 km from the maximum voltage in the thundercloud, the TGEs can be unleashed. Therefore, we can expect that TGEs detected on Aragats highland extended to many km$^2$, and in whole huge space above RREAs are producing billions of electrons, gamma rays, and neutrons. The experimental illustration of this claim is a perfect coincidence of the time-series of particle count rates measured on Aragats by the remote detectors (at distance ≈300 m) and measured in Japan 2 gamma glows located 1.35 km apart [23].

**Table 1. Characteristics of five TGEs terminated by lightning flash during storm occurred on 9 June 2021**

| # | TGE Duration | Lightning type and time | Distance to lightning flash (km) | NS electric field(kV/m) | TGE enhancement (%) |
|---|---|---|---|---|---|
| | | | | | |
| TGE1 | 14:48:51 – 14:50:27 36 s | Inverted-polarity IC followed by -CG 14:50:27.912 | 2.4 | +13 | 3.8 |
| TGE2 | 14:51:06- 14:52:14 1 min 8 sec | -CG 14:52:14.510 | 1.6 | -15 - 9 | 8.3 |
| TGE3 | 14:52:38 – 14:53:31 53 s | Inverted-polarity IC 14:53:31.443 | 1.6 | -10 +16 | 3.5 |
| TGE4 | 14:53:47- 14:54:31 44sec | No nearby flash detected | - | 16 | 3.1 |
| TGE5 | 14:57:07 – 14:59:02 1 min 55 sec | -CG 14:59:01.705 | 7.5 | -8 | 5.2 |

In Fig. 2 we show the energy release histograms measured by the ASNT spectrometer. The sequence of the 20-sec histograms of the energy releases in the 60 cm thick scintillator are denoted by different colors. The large area of spectrometer provides the possibility of reliable recovering of the energy spectra, even for short TGEs, see Fig. 3 where we present energy spectra of 4 TGEs from 5. In Fig.2 we see that after each of 4 lightning flashes (the time of flash is shown in red boxes) the particle flux abruptly terminated to restart after several minutes (or tens of seconds). Thus, the lightning flash does not fully quench a huge voltage in the thundercloud, the charging machine during the active phase of the storm continues to separate charges and generate an electric field above the critical value to unleash the runaway process again and again.

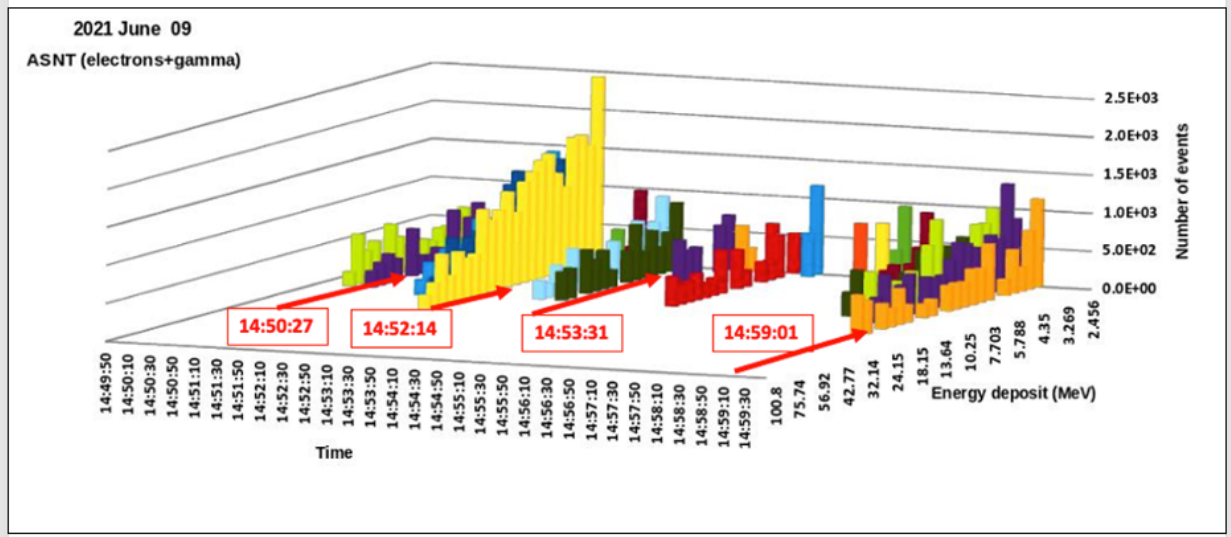

**Figure 2.** The 20-sec histograms of the energy releases in the 60 cm thick scintillator of the ASNT spectrometer. By red arrow and boxes, we denote the time of the lightning stroke that abruptly terminates TGE.

The recovered energy spectra of all TGEs (see details of spectra recovering in [9]) extended well above 4 MeV, thus, possible contamination of spectra by Radon progeny gamma radiation can be neglected and we can confirm that registered enhancements are due to RREA processes in the atmosphere, which were terminated by a nearby lightning flash. As the accelerating electric field declined well above the earth's surface electrons lost all energy to ionization and do not reach the ground. Thus, there was no possibility to recover the electron energy spectrum. A small amount of Compton scattered electrons (<2% of gamma ray flux) generated by the gamma ray "beam" in the atmosphere above detectors dissipate in the background noise.

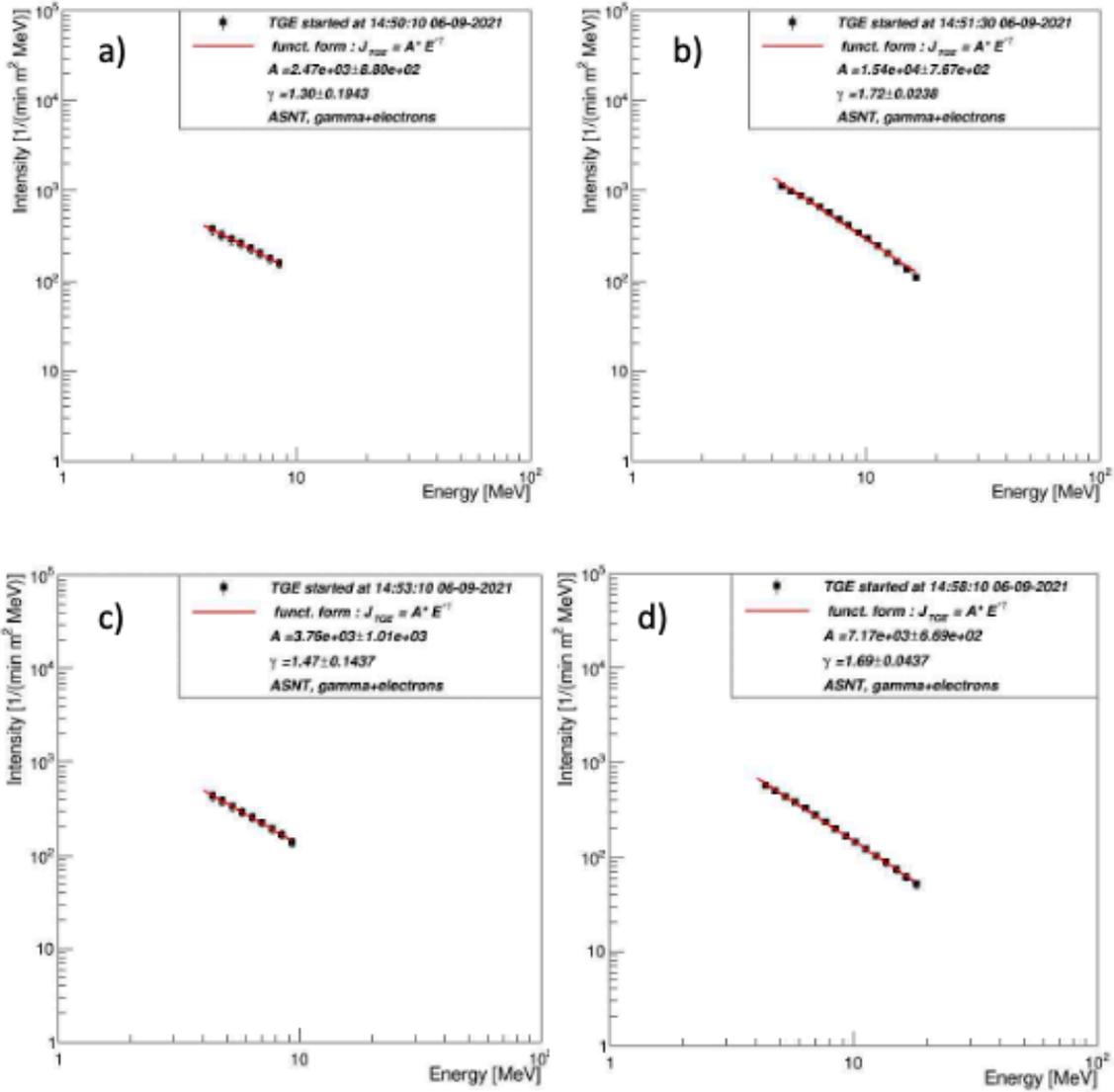

*Figure 3.* **Recovered differential energy spectra of the TGE particles, registered by ASNT detector (20 sec time series of histograms of energy releases in 60 cm thick scintillator).**

Thus, if the atmospheric electric field is stronger than a critical one (specific for particular air density), free electrons from the EASs with energies corresponding to minimum ionization losses (0.1-2 MeV) get energy from the electric field, run away, knock-on electrons from atoms and initiate electron-gamma ray avalanches. If the electric field prolongs to low enough above the earth's surface, gamma rays and electrons, can reach the earth's surface and be registered as TGE. At some not yet clear exceptional conditions – gamma rays can reach orbiting gamma ray observatories and be registered as TGFs. Relatively long TGEs are simply a set of consequent avalanches (or microbursts [24]) integrated in time. Thus, TGFs and TGEs share the same origin, namely the runaway process using EAS electrons as seeds. However, only very few most energetic gamma rays occasionally reach the orbiting gamma detectors. ASIM detector located on the Space station and specially constructed for the TGF detection already measured several gamma rays one following another [25].

When electrons ionized the atmosphere and enable a path for the lightning leader, a lightning flash abruptly terminates the RREA and consequently the TGE. If the electric field is maximum not in the vicinity of particle detectors, but more than 10 km away, the remote lightning flash will not abruptly reduce the electric field and TGE will finish smoothly. However, the location of Aragats station nearby the south summit of Aragats, makes frequent the first scenario of the TGE interruption, as one can see in Fig. 1 and in figures of 165 TGE events collected during the last 5 years (see Mendeley data V1, doi: 10.17632/p25bb7jrfp.1). For each of 165 observed TGEs we show the time series of particle detector count rates and corresponding disturbances of the near-surface electric field at Aragats and Nor Amberd stations (3200 and 2000 m above sea level, 12 km apart. All TGE events were stopped by lightning flashes; at initiation and mature (maximum flux) of the TGE development no lightning activity was registered. Thus, a vast amount of the collected data demonstrates that particle fluxes are not originated by the lightning flashes, the lightning flashes terminated particle fluxes by the lowering (quenching) potential difference in the atmosphere.

### 3. Neutron Monitors as EAS burst detectors

The Aragats neutron monitor (ArNM) consists of 18 cylindrical proportional counters of CHM-15 type (length 200cm, diameter 15cm) filled with $BF_3$ gas enriched with $B^{10}$ isotope. The proportional chambers are surrounded by 5 cm of lead (producer) and 2 cm of polyethylene (moderator). The cross-section of the lead producer above each section has a surface area of $6m^2$, and the total surface area of the three sections is 18 $m^2$. The atmospheric hadrons and gamma rays produce secondary neutrons in lead absorber. Then, the neutrons were thermalized in a moderator, enter the sensitive volume of the counter, and yield $Li^7$ and α particles via interactions with boron gas. The α particle accelerates in the high electrical field inside the chamber and produces a pulse registered by the data acquisition electronics. The high-energy hadrons generate a large number of secondary neutrons entering the sensitive volume of the proportional counter, and if all pulses initiated by the incident hadrons need to be counted, the dead time of the NM should be maintained very low (the ArNM has a minimal dead time of 0.4μs). If only the incident hadrons need to be counted (a one-to-one relationship between count rate and hadron flux), the dead time must be as long as the whole secondary neutron collection time (1250 μs), to avoid double-counting. In [26] was described detection of the neutron bursts in the NM related to occasional hitting of the detector by a core of high-energy EAS. EAS hadrons and gamma rays will generate numerous thermal neutrons in the construction elements of the building and in the soil, enormously increasing the NM count rate (size of the neutron burst or peak multiplicity). This option of EAS core detection by NM was almost not recognized in the past, because the usually used long dead time (1250 μs), does not permit counting of the neutron multiplicity. With establishing 3000 times shorter dead time and 1-second collecting time for ArNM (usually only 1-min time series are available for NM), we detect EASs hitting NM, several of which provide bursts with a number of particles exceeding 2000 (see Figs 20-22 in [27]). The neutron burst sizes above 2000 are extremely rare (one to two per month. The primary particle energies corresponding to these events are very high (>10 PeV).

In 3 years, we have detected only two strong negative CG lightning flashes coinciding with large enhancements in neutron monitor. However, these 2 events were detected on the maximum phase of TGE, which was abruptly terminated. Thus, the lightning flash stopped particle burst, not initiate it (see Figs. 16 and 17 of [27]).

An example of particle bursts produced by EAS is demonstrated in Figs. 4 and 5. The burst was detected on a non-thunderstorm day, without any relation to lightning flash.

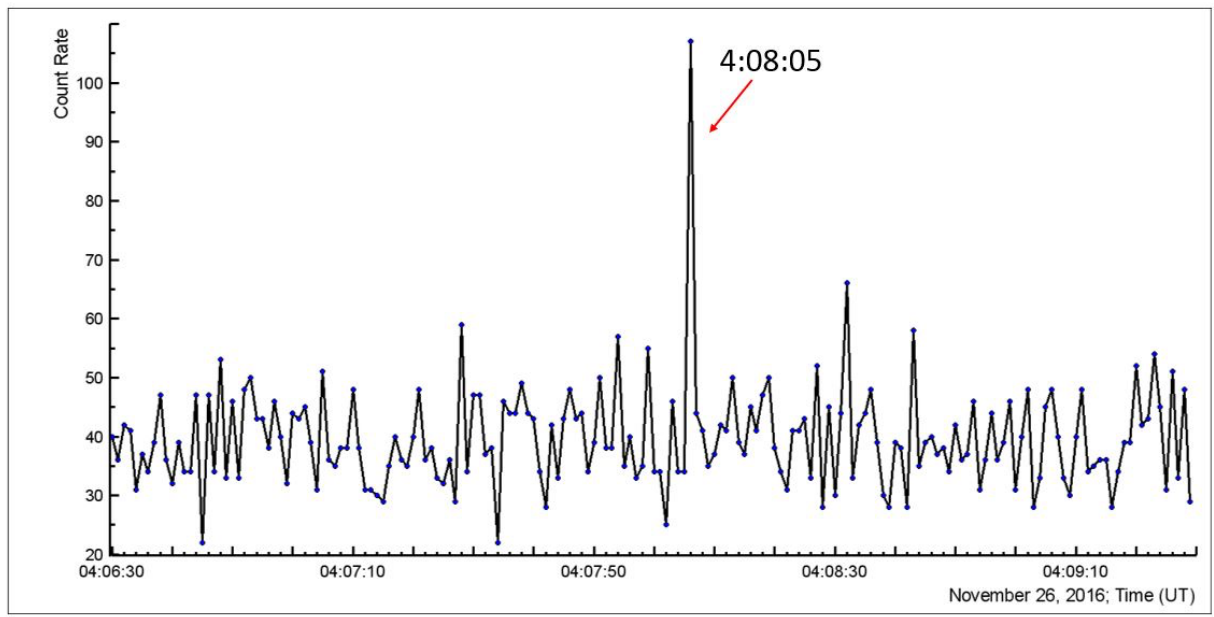

**Figure 4. The 1-sec time series of the count rate of the Aragats Neutron Monitor (the proportional chamber N2). A neutron burst with multiplicity of 107 is registered at 4:08:05 on November 26, 2016.**

The sequence of pulses from the proportional counter of ArNM was recorded by a Picoscope 5244B digitizing oscilloscope. The record length was 100 ms including 20 ms pre-trigger time and 80 ms post-trigger time. The sampling rate was 250 MS/s (sampling interval of 4 ns), and the amplitude resolution was 8 bits. The signal of the Aragats Neutron Monitor was also relayed to the MyRIO board (National Instruments) which produced a pulse for the oscilloscope triggering when the count rate of the detector exceeds a preset threshold value (usually a 20% enhancement relating the running mean). The oscilloscope records of the neutron burst are shown in Fig.5. The observed burst is rather "dense" in the beginning (interval between pulses is few microseconds) and much sparser at the end (interval between pulses is from tens to hundreds of microseconds). Most frequently, the pulse amplitude is the largest in the beginning of the burst. Interval between pulses varies during the burst: it is the shortest (about 3-5 µs) in the beginning, and increases to tens and hundreds of microseconds at the end of the burst. All analyzed 50 bursts were observed on non-thunderstorm days when the near-surface electric field didn't exceed 0.4 kV/m. Bursts were observed as sequences of microsecond pulses temporally isolated from other pulses on a time scale of at least 100 microseconds. The burst duration is defined here as a time interval between the first and last detectable pulses in the sequence. Distribution of burst durations for 50 observed events is shown in Fig.6.

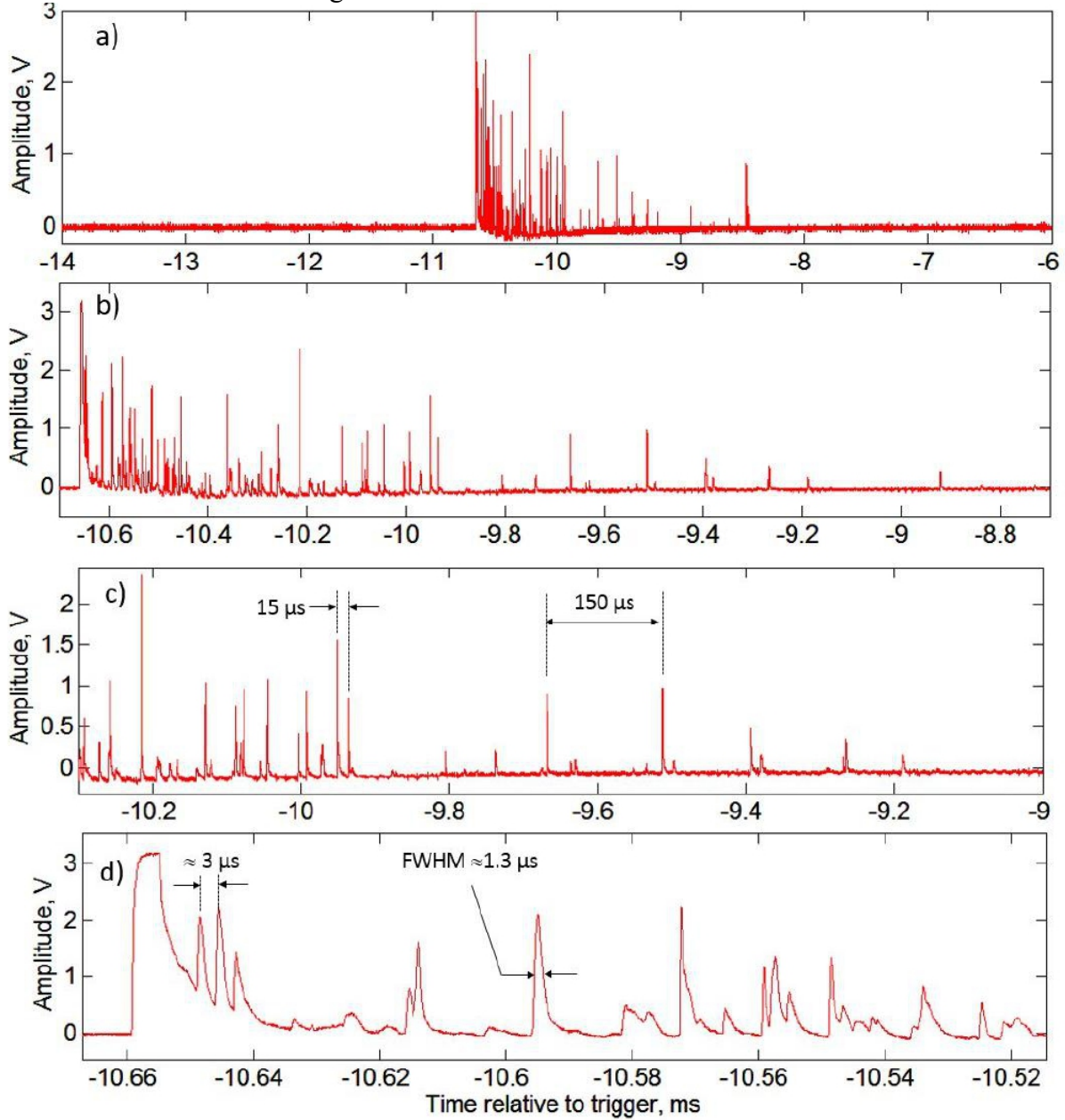

**Figure 5. Oscilloscope records of neutron burst that occurred at 4:08:05 on November 26, 2016. Burst duration is ≈ 2.2 ms, the multiplicity is 107 per m$^2$. The four panels (a-c) show the records of the burst on different time scales.**

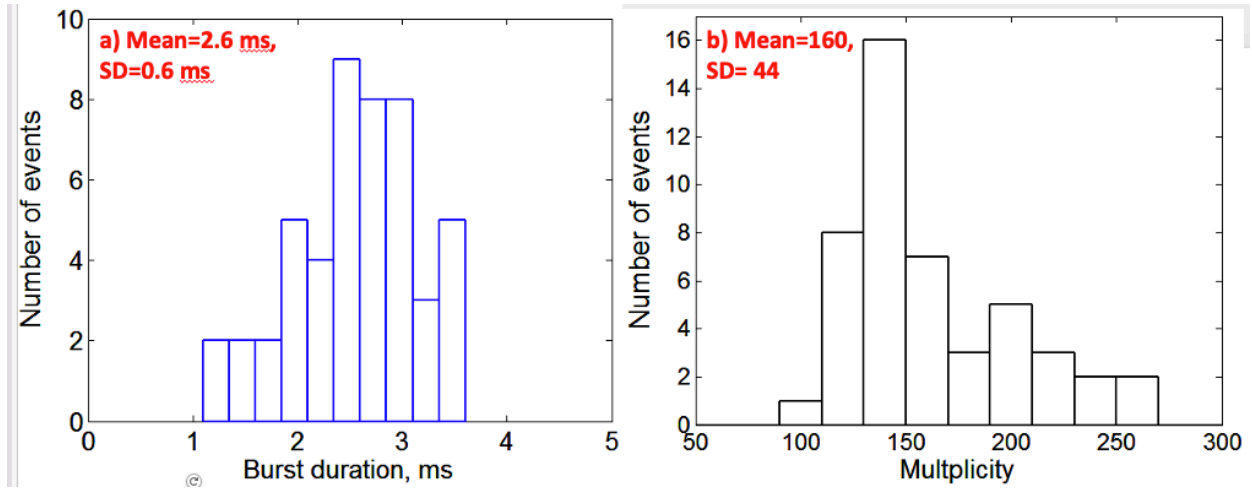

**Figure 6. Histogram of the neutron burst duration (a) and corresponding multiplicity histogram (b).**

As we can see in Fig. 6 neutron monitor is enlarging very short EAS time profile (20 – 30 ns) by ≈ 5 orders of magnitude (2.6 ms) making it possible to use rather a slow device (neutron monitor) for registration of EAS and, after detailed simulation of the detector response function, to estimate primary particle energy. In Fig. 23 of [27], we show the energy spectra of primary cosmic radiation obtained by the relation of the frequency of different observed multiplicities (neutron burst sizes) to integral energy spectrum measured by MAKET array in the energy range near the "knee" of all particle spectra – 3-5 PeV [28]. An abundant number of the EAS particles that concentrated near the shower core highly saturate the array detectors (usually plastic scintillators), making it impossible to research the hadron distribution in the EAS cores. Due to this obstacle, the linear distribution function of the EAS electrons (used to recover the shower size and then – energy) is approximated from a distance of 5 m of the shower core. Thus, by registering neutron multiplicities the distribution of the parent hadrons at the EAS core also can be recovered and studied.

In the next section we will demonstrate that small particle detectors operated within large surface arrays routinely register particle bursts coinciding with EAS observations.

4. **Possible scenarios of DTGF origination**

High Altitude Water Cherenkov Array (HAWC, [21]) consisted of 300 water Cherenkov detectors of 4 meters in height and 7.3 meters in diameter. A small, fast scintillator detector (7.62 × ⌀7.62 cm LaBr3) located nearby water-Cherenkov detectors, occasionally observed large bursts followed by a significant number of delayed counts. The small detector output was attached to the Broadband Interferometric Mapping and Polarization (BIMAP) sensor's electronics [29]. For each trigger, BIMAP captures 15 ms of data with 5 ms pre-trigger data [30].

All neutron bursts observed at HAWC between September 2017 and September 2019 occurred during *fair-weather days*, meaning that there was no nearby thunderstorm activity, see Tab. 1 of [31]. CORSIKA [32] simulations confirm that observed bursts were originated by neutrons from large cosmic-ray showers captured in nuclei of soil which produced high-energy gamma-rays through (n, γ) reactions. Thus, observed by HAWC particle bursts, as bursts detected on Aragats were not related to atmospheric discharges. The observed bursts are initiated by particles belonging to EASs cores with energies above 10 PeV hitting the HAWC array.

Another large detector, the Telescope Array (TA) is composed of 507 scintillator detectors on a 1.2-km square grid occupying totally 700-km$^2$ area (TASD array, [22]). The surface array provides shower footprint information including core location, lateral density profile, and timing, which are used to reconstruct shower axes and energy. Each measuring unit consists of upper and lower scintillators of 1 cm thickness and 3 m$^2$ area. The upper and lower planes are separated by a 1-mm-thick steel plate and are readout by photomultiplier tubes that are coupled to the scintillator via an array of wavelength-shifting fibers. The output signals from the photomultipliers are digitized by a 12-bit ADC with a 50-MHz sampling rate. An event trigger (frequency ≈0.01 Hz) is recorded when three adjacent units observe a signal larger than three vertical equivalent muons (VEMs, ≈ 2 MeV per 1 cm of scintillator) within 8 $\mu$s.. When a trigger occurs, the signals from all units within ±32 $\mu$s, which detect an integrated amplitude greater than 0.3 VEM are also recorded. The efficiency of registering high-energy photons on average is proportional to the thickness of the scintillator in cm (1-2% for the 1 cm thick scintillator).

The bursts of consecutive TASD triggers were recorded in 1-ms time intervals in correlation with lightning flashes above the TA detector, observed by the lightning mapping array (LMA) and the Vaisala National Lightning Detector Network (NLDN). The Lightning flashes that produce trigger bursts were very rare. There are typically about 750 NLDN-recorded flashes (IC and cloud to ground) per year over the 700-km$^2$ TASD array. In 8 years of TA operation, there were identified only 20 bursts correlated with lightning activity [33]. Thus, fewer than 0.5% of NLDN flashes recorded over the TASD were accompanied by identifiable gamma bursts; that is approximately the fraction of thunderstorms to fair weather duration. Authors of [34] conclude that the observed TASD bursts are due to primary gamma-ray showers with a fluence of ≃$10^{12}$-$10^{14}$ photons with energies above 1 MeV. The burst durations were found to be within several hundred microseconds, the altitude of the source was typical of a few km above ground level. However, the authors do not relate these showers to EASs, as in the HAWC and Aragats experiments, but to a downward negative leader, which ends up in a -CG discharge. In the recent publication [35] they introduce a lightning-related scenario of the "downward TGF" origination: "The results show that the TGFs occur during strong initial breakdown pulses (IBPs) in the first few milliseconds of negative cloud-to-ground and low-altitude intracloud flashes, and that the IBPs are produced by a newly-identified streamer-based discharge process called fast negative breakdown." And "In addition to showing the nature of IBPs and their enigmatic sub-pulses, the observations also provide a possible explanation for the unsolved question of how the streamer to leader transition occurs during the initial negative breakdown, namely as a result of strong currents flowing in the final stage of successive IBPs, extending backward through both the IBP itself and the negative streamer breakdown preceding the IBP." Thus, to explain HAWK and Aragats bursts authors use well-known EAS physics, and for an explanation of the analogical bursts observed in the TA experiment – a streamer to leader transition including IBPs and their enigmatic sub-pulses.

It is strange, that TA does not report any bursts at fair weather that should hit TA array as frequently as HAWK and Aragats. In the HAWC and Aragats experiments, the particle bursts and lightning activity are completely separated. EAS propagation model satisfactorily explains registered at fair weather particle bursts. No relation to lightning activity was assumed and observed. Possibly, the particle bursts and discharge processes registered by the TA array are not causally related, but randomly coincide only? As was discussed above, fewer than 0.5% of total lightning flashes above TA produce bursts. Large EASs are more-or-less uniformly distributed in time, not influenced by weather at all. Thus, maybe the TA burst events are a subsample of the EASs triggers that occasionally coincide with lightning activity?

Recently another scenario of the origination of intense burst was suggested in [36]. The point corona discharges usually occurred beneath thunderstorms very near to the tips of the grounded sharp conductors when electric fields at the ground reach threshold values of ~3 to 5 kV/m. In a small discharge region, the strength of the electric field can reach 2 MV/m, and EAS-originated electrons will run away and make electron-photon avalanches producing an intense burst of particles on the earth's surface. However, because of the stochastic nature of the corona discharges, these small local regions will emerge spontaneously and not coherently. Thus, the TGEs observed by the remote detectors (and detectors inside the buildings) cannot demonstrate the coherent and smooth enhancement and decay of particle flux.

Another exotic hypothesis to explain the particle flux enhancements during thunderstorms is the ball lightning origination in the skies with consequent intense radiation of gamma rays [37]. This hypothesis identifies multiple light spots observed during TGEs in the skies above Aragats [38] with a ball lightning system emerging in the electrified atmosphere. However, as for the previous hypothesis, the emergence of the stable gamma ray flux registered by detectors covering many thousands of $m^2$ should be explained.

**Conclusions**

Enhanced particle fluxes emerging in space and on the earth's surface during thunderstorms are produced in the atmospheric electric field by the runaway process when free electrons from small and large EASs enter an electric field, which strength is larger than the critical value.

Analysis of the particle burst observational data from HAWC and Aragats experiments allows us to conclude that the measured bursts are not causally related to the lightning flashes, but can be explained by a common EAS physics only.

Thus, EAS physics and HEPA are synergistically connected and need to exchange results for the explanation of particle bursts and for revealing the influence of atmospheric electric fields on the EAS shape and size.

The "downward TGFs" reported by the TA collaboration can be interpreted as high-energy EASs occasionally hitting the detector area, without any relation to atmospheric discharges. To prove the lightning nature of the bursts it is necessary to collect much more frequent bursts at fair weather and compare them with burst data collected during thunderstorms.

To prove the corona discharge and ball lightning scenarios authors should demonstrate that the corona discharge at multiple not connected metallic structures and emitting by randomly emerged ball lightning system in the skies can produce a uniform electric field above the ground on thousands of square meters area, and that continuous discharges can sustain such a uniform field for minutes.

The network of near 50 Neutron monitors operate at different altitudes, latitudes, and longitudes for more than 60 years [39]. Maintenance of such a detector is very cheap and they are providing data for many years with minimal intervention of personal. The data stream is collected in the databases with open access and a user-friendly interface [40]. By using the neutron monitor database (NMDB) after a very simple modernization of NM electronics, it will be possible to recover energy spectra of galactic cosmic rays all around the globe, as we demonstrated in [27].

The described method (registration of the neutron multiplicity) is used at the Scientific and Educational Center NEVOD (MEPhI) to detect the neutron and charged components of EAS at primary particles' energies around and above the "knee". The mean time of the neutron "bursts" measured by the URAN array (see Fig 3 in [41] well coincides with the measured by the ArNM value (Fig.6). The neutron multiplicity method also is used for the investigation of hadrons in the cores of high-energy EAS at the Tien Shan high-altitude cosmic ray station [42].

 Acknowledgments
We thank the staff of the Aragats Space Environmental Center for the uninterruptible operation of experimental facilities on Aragats *under severe weather conditions*. The authors thank S. Soghomonyan for providing information about lightning flashes on 9 June 2021, for analysis of the bursts registered by neutron monitor, and for preparing the Mendeley Dataset. Special thanks to Yuri Stenkin, who first explain long-living EAS phenomena. The data for this study is available in numerical and graphical formats by the multivariate visualization software platform ADEI on the WEB page of the Cosmic Ray Division (CRD) of the Yerevan Physics Institute, http://adei.crd.yerphi.am/adei. In the WIKI section explanations of facilities and how to select necessary data are posted. *The authors acknowledge the support of the Science Committee of the Republic of Armenia, in the frames of the research project № 21AG-1C012.*

**Declaration of Competing Interest**

The authors declare no conflict of interest.

**Data Availability Statement**

The data for this study are available in numerical and graphical formats on the WEB page of the Cosmic Ray Division (CRD) of the Yerevan Physics Institute, http://adei.crd.yerphi.am/adei and from Mendeley datacets [43-45].

[44] Soghomonyan, Suren; Chilingarian, Ashot (2021), "Thunderstorm ground enhancements abruptly terminated by a lightning flash registered both by WWLLN and local network of EFM-100 electric mills.", Mendeley Data, V1, doi: 10.17632/ygvjzdx3w3.1

[45] Chilingarian, Ashot, Hovsepyan, Gagik,
Dataset for 16 parameters of ten thunderstorm ground enhancements (TGEs) allowing recovery of electron energy spectra and estimation the structure of the electric field above earth's surface, Mendeley Data, V1,
doi: 10.17632/tvbn6wdf85.2